\documentclass[12pt]{article}
\usepackage{amsmath, amssymb, graphicx, hyperref, braket, xcolor, booktabs}
\usepackage[margin=1in]{geometry}

\newtheorem{proposition}{Proposition}

\newtheorem{remark}{Remark}

\title{A Periodic Orbit Trace Formula for Quantum Scrambling: \\
The Role of the Normally Hyperbolic Invariant Manifold%
\footnote{The result is a semiclassical expansion; the term ``trace formula'' 
is used in the sense of a formal weighted periodic‑orbit sum.}}
\author{Stephen Wiggins \\ 
\small Hetao Institute of Mathematics and Interdisciplinary Sciences, 
Shenzhen, Guangdong Province, China \\ 
\small School of Mathematics, University of Bristol, 
Bristol, BS8 1UG, United Kingdom}
\date{\today}

\begin{document}

\maketitle

\begin{abstract}
Out-of-Time-Order Correlators (OTOCs) quantify quantum information scrambling,
but their connection to localized phase-space structures—such as chemical 
transition states—has remained largely unexplored. We derive a formal 
leading-order semiclassical expansion for the microcanonical OTOC in systems 
with an index-1 saddle point, expressing the scrambling rate as a coherent sum 
over unstable periodic orbits on the Normally Hyperbolic Invariant Manifold 
(NHIM). Valid in the semiclassical limit and the intermediate-time regime 
before the Ehrenfest time, our derivation utilizes the Normal Form theory of 
the transition state, which transforms the Hamiltonian near the saddle into an 
integrable (though generally non-separable) form dependent on conserved actions. 
We outline the derivation of the microcanonical trace, the semiclassical 
propagator for integrable systems, the factorization of the stability matrix, 
and the explicit Schur complement reduction of the stationary phase approximation. 
Our result extends periodic-orbit trace methods to scrambling observables, 
recovering a local instability exponent $\Lambda(J)$ governing the leading 
semiclassical growth window. As a special case, when the observation time 
coincides with the intrinsic periods of the contributing orbits, the trace 
sum reduces to an effective $1.5\Lambda$ scaling, reflecting the competition 
between local hyperbolic growth and wavepacket dilution. We emphasize that 
this simplified form is conditional; the full expansion retains a coherent 
sum over orbit periods. Finally, we discuss how the dependence 
of the instability on transverse actions suggests a mechanism for mode-selective 
control of scrambling, and outline a numerical evaluation strategy to test these 
predictions.
\end{abstract}

\textbf{PhySH Concepts:} Quantum Chaos, Semiclassical Methods, Reaction 
Dynamics, Information Scrambling

\tableofcontents

\section{Introduction}

\subsection{OTOCs, scrambling, and classical instability}
Out-of-time-order correlators (OTOCs) have emerged as central tools for
characterizing the growth of operator complexity and the spread of quantum
information under unitary dynamics. Originally introduced by Larkin and
Ovchinnikov in their study of superconductivity in disordered systems
\cite{Larkin1969}, the OTOC has been rediscovered in the context of quantum
chaos, many-body dynamics, and black hole physics \cite{Maldacena2016}. A
commonly used definition is the squared commutator
\begin{equation}
C(t) = \langle [\hat{W}(t), \hat{V}(0)]^\dagger [\hat{W}(t), \hat{V}(0)] \rangle
\end{equation}
which quantifies the degree to which initially commuting operators fail to
commute at later times. This growth reflects the spreading of initially
localized quantum information into increasingly complex degrees of freedom,
a process known as scrambling \cite{Sekino2008}.

In semiclassical systems, scrambling is closely related to classical dynamical
instability \cite{Rozenbaum2017,Jalabert2018}. Using semiclassical
propagation and Weyl symbol correspondence (where a quantum operator is
mathematically mapped to its equivalent classical phase-space function), the
leading symbol of the commutator is governed by the classical stability
matrix. This leads to exponential growth of the OTOC in systems with hyperbolic
dynamics, with growth rates determined by classical instability exponents
\cite{Rozenbaum2017,Jalabert2018,Rammensee2018}. Semiclassical expansions of OTOCs in
low-dimensional chaotic systems have explicitly demonstrated this
Lyapunov-governed exponential window and the transition to
interference-dominated saturation \cite{Rozenbaum2017,Jalabert2018,Rammensee2018}.

However, scrambling is fundamentally an operator-level quantum phenomenon and
cannot be reduced entirely to global classical chaos. Crucially, recent work
by Xu, Scaffidi, and Cao has demonstrated that the parametric exponential
growth of OTOCs does not necessitate global chaos \cite{Xu2020}. Instead,
they revealed that scrambling can simply result from the presence of isolated,
unstable fixed points in phase space, even if the surrounding system is
classically integrable. Direct studies of exponential OTOC growth in
inverted-harmonic-oscillator potentials have further confirmed that scrambling
can occur without thermalization \cite{Hashimoto2017,Kidd2021}, alongside
related saddle-dominated operator-growth frameworks \cite{Bhattacharjee2022}.
This insight provides the foundational motivation for our present
study: the quantum butterfly effect can be rigorously analyzed by isolating
the dynamics around an index-1 saddle point.

\subsection{Scrambling in Chemical Reactions and the NHIM}
The realization that saddle points can independently drive information
scrambling has direct implications for chemical physics. A chemical
reaction crossing a potential energy barrier is, dynamically, a traversal of
an index-1 saddle in phase space. Recent seminal studies by Zhang et al. have
pioneered the application of OTOCs to model chemical reactions, demonstrating
how the quantum Lyapunov exponent approaches fundamental scrambling bounds in
the tunneling regime \cite{Zhang2024}. Parallel investigations by Sadhasivam
et al. have shown that instantons—delocalized structures dominating tunneling
statistics—and quantum thermal fluctuations tightly control this quantum bound
to chaos \cite{Sadhasivam2023}, a connection further illuminated by the
commentary of Wolynes and Gruebele \cite{Wolynes2023}. Furthermore, the
exponential operator growth in Hilbert space has recently been tied directly
to the off-diagonal decay of operator matrix elements in the system eigenbasis
\cite{Sadhasivam2025}.

To rigorously map these scrambling phenomena to classical trajectories, we
rely on the phase-space geometry of the transition state. The phase-space
structure near such saddle points is organized by a Normally Hyperbolic
Invariant Manifold (NHIM), which acts as the dynamical core of the transition
state. The NHIM provides a rigorous phase-space framework for transition state
theory and organizes reactive trajectories \cite{Wiggins1994}. Near the saddle,
the Hamiltonian admits a normal-form representation in which the dynamics form
an integrable structure composed of a hyperbolic reaction coordinate and stable
bath degrees of freedom \cite{Waalkens2008}. To ensure the formal integrability
of this normal form to arbitrary orders, we strictly assume that the linear
bath frequencies are non-resonant. The presence of resonances would require a
resonant normal form that retains some angular couplings, complicating the
invariant torus structure necessary for the subsequent trace evaluation. Under
the non-resonant assumption, this normal-form structure provides a natural
coordinate system for analyzing semiclassical propagation and instability.

\subsection{Semiclassical trace formulas and NHIM-based reaction dynamics}
Semiclassical trace formulas provide a fundamental link between classical
dynamics and quantum observables. Gutzwiller's trace formula expresses the
density of states of chaotic systems as a sum over unstable periodic orbits
\cite{Gutzwiller1971}, while Berry and Tabor derived an analogous formula for
integrable systems organized by resonant invariant tori \cite{Berry1976}. As
Richter, Urbina, and Tomsovic have recently detailed, these periodic orbit
frameworks form the semiclassical roots of universality in many-body quantum
chaos \cite{Richter2022}.

These methods have been extended to open systems and transition-state dynamics.
In particular, Schubert, Waalkens, Goussev, and Wiggins derived periodic-orbit
trace formulas for quantum reaction rates \cite{Schubert2010}. Their approach
successfully unified the hyperbolic flux across the saddle with the Berry-Tabor
sum over the stable NHIM tori. Our derivation leverages this identical hybrid
phase-space geometry, but we replace the reaction flux operator with the OTOC
squared-commutator insertion. This shifts the semiclassical focus from mass
transport across the barrier to the spreading of operator complexity within
the saddle neighborhood.

\subsection{Relation to trace formulas for inserted observables and regularizations}
Our derivation effectively aims to construct a trace formula featuring an
operator insertion (the scrambling weight $\hat{M}^\dagger \hat{M}$). This
positions our work within the broader mathematical physics literature of trace
formulas, where we distinguish three broad categories of complexity:

\begin{enumerate}
    \item \textbf{Classic spectral traces:} Density-of-states formulas such as
    Gutzwiller's and Berry-Tabor's. These formulas evaluate the trace of the
    bare quantum propagator $e^{-i\hat{H}t/\hbar}$ (or the energy-dependent
    resolvent) to strictly isolate the energy eigenvalues, without inserting
    any additional observable operators into the integral.
    \item \textbf{Weighted trace formulas:} Traces incorporating smooth, static
    observables, often rigorously validated via coherent state decompositions
    \cite{Combescure1999}.
    \item \textbf{Dynamical insertions (The present case):} The inserted
    observable is a squared commutator whose Weyl symbol—the classical
    phase-space function representing the operator—grows exponentially with
    the observation time $t_{OTOC}$. This parametric growth is atypical for
    ``smooth static observable'' trace contexts and fundamentally changes the
    stationary phase evaluation.
\end{enumerate}

Furthermore, the choice of how the OTOC is thermally weighted—a process known
as regularization—is critical \cite{RomeroBermudez2019}. Standard thermal
formulations often split the Boltzmann operator $e^{-\beta \hat{H}}$ symmetrically
around the commutators to facilitate analytical tractability. However, this
global thermal smoothing can artificially redistribute the observed exponential
weights in finite systems and obscure the underlying local dynamics. To
rigorously isolate the specific phase-space geometry of the saddle without
these global thermal averaging effects, we focus on an unnormalized, purely
microcanonical projection at a fixed energy $E$ \cite{Hashimoto2017}.

\subsection{Contributions and Main Results}
The specific contributions of this work are organized into three primary
components:

\begin{enumerate}
    \item \textbf{Leading-order semiclassical expansion:} We derive a formal
    semiclassical expansion for the microcanonical OTOC $C_{E}(t_{OTOC})$ as a
    coherent sum over resonant tori on the NHIM, keeping the spectral period
    ($\tau_\gamma$) and observation time ($t_{OTOC}$) strictly separated. This
    result is a weighted periodic-orbit sum in the sense that the squared
    commutator contributes an orbit-dependent factor $e^{2\Lambda_\gamma t_{OTOC}}$.
    \item \textbf{Interpretive special case:} When the observation time coincides
    with the intrinsic periods of the dominant periodic orbits (a non‑generic
    resonance condition), the trace sum simplifies to an effective $1.5\Lambda$
    scaling. We explicitly note that this is a special case, not a generic
    prediction.
    \item \textbf{Application and future validation:} Using an Eckart-Morse
    normal form, we explicitly construct the interference sum and outline 
    quantitative benchmarks for a numerical evaluation. The normal‑form 
    coefficients are taken from the verified 10th‑order truncation of Waalkens, 
    Schubert, and Wiggins \cite{Waalkens2008}.
\end{enumerate}

\section{The Dynamical Framework: Normal Forms on the NHIM}

The transition from abstract operator scrambling to a computable semiclassical
trace formula requires a precise geometric framework. In a multidimensional
system, the hyperbolic instability responsible for scrambling is initially
intertwined with the surrounding stable vibrational modes. Evaluating a quantum
path integral directly in these physical coordinates is analytically intractable
due to these nonlinear couplings.

The objective of this section is to construct a canonical coordinate system
that mathematically isolates the saddle's hyperbolic instability from the stable
background, while preserving the exact nonlinear interactions that dictate the
local scrambling rate. We achieve this by transforming the local dynamics into
an integrable representation organized around a Normally Hyperbolic Invariant
Manifold (NHIM). 

We proceed in five steps. First, we isolate the fundamental hyperbolic geometry
of the reaction coordinate using canonical scaled variables (Section 2.1),
before embedding it within the full multidimensional phase space (Section 2.2).
To handle the energetic couplings, we apply the Poincaré-Birkhoff Normal Form
transformation (Section 2.3), which provides the critical step of integrating
the local dynamics into a function of conserved actions. This framework allows
us to rigorously define the NHIM and extract the nonlinear frequencies and
instability exponents that govern the phase space (Section 2.4). Finally, we
delineate the specific semiclassical and temporal boundaries within which this
normal-form approximation remains valid (Section 2.5). By the end of this
section, we will have established the exact classical variables—and their
specific block-triangular stability properties—required to evaluate the
semiclassical trace integral in Section 3.

\subsection{Canonical Scaling of the Reaction Coordinate}
Before analyzing the full high-dimensional transition state, it is instructive
to isolate the fundamental dynamical structure responsible for scrambling: the
pure hyperbolic instability along the reaction coordinate. In the vicinity of
any index-1 saddle, the motion transverse to the barrier can be locally
approximated by a one-dimensional inverted harmonic oscillator. For the
quadratic saddle Hamiltonian,
\begin{equation}
H_{reac} = \frac{p_u^2}{2m} - \frac{m\lambda^2}{2}q_u^2
\end{equation}
the linearized flow yields the monodromy matrix:
\begin{equation}
M_{reac}(t) = \begin{pmatrix} \cosh(\lambda t) & \frac{1}{m\lambda}\sinh(\lambda t) \\ m\lambda \sinh(\lambda t) & \cosh(\lambda t) \end{pmatrix}
\end{equation}

To reconcile the physical parameters with the symmetric matrix forms utilized
in subsequent trace evaluations, we apply a canonical symplectic scaling
transformation $(q, p) \mapsto (q\sqrt{m\lambda}, p/\sqrt{m\lambda})$. In these
scaled canonical coordinates, the effective mass drops out, the Hamiltonian
becomes $H_{reac} = \frac{\lambda}{2}(p_u^2 - q_u^2)$, and the monodromy matrix
becomes perfectly symmetric:
\begin{equation}
M_{reac}(t) = \begin{pmatrix} \cosh(\lambda t) & \sinh(\lambda t) \\ \sinh(\lambda t) & \cosh(\lambda t) \end{pmatrix}
\end{equation}

This idealized $2 \times 2$ matrix serves as the central mathematical object
of our derivation. As we will show in Section 3, when the full system is
transformed into Normal Form, $M_{reac}(t)$ explicitly becomes the unstable
diagonal block of the complete high-dimensional Monodromy matrix.

\subsection{Phase Space Geometry of the Transition State}
We consider a Hamiltonian system with $f + 1$ degrees of freedom. The phase
space is the symplectic manifold $\mathbb{R}^{2(f+1)}$ equipped with the standard
symplectic form. We focus on the dynamics in the neighborhood of an index-1
saddle point (equilibrium), which governs the crossing of a potential barrier.
This equilibrium point is characterized by one unstable degree of freedom (the
reaction coordinate) and $f$ stable degrees of freedom (the bath modes). In the
linearized approximation, the eigenvalues of the Hamiltonian vector field at the
saddle are $\pm\lambda$ (associated with the unstable direction) and $\pm i\omega_k$
for $k = 1, ..., f$ (associated with the stable directions). 

To guarantee the validity of the normal form transformation that follows, we 
must impose a strict non-resonant condition on these stable frequencies. We 
assume that the linear bath frequencies $\omega_k$ are rationally independent, 
meaning there exists no integer vector $\mathbf{k} \in \mathbb{Z}^f \setminus \{0\}$ 
such that $\sum_{j=1}^f k_j \omega_j = 0$. This non-resonant geometric 
structure forms the rigorous basis for the Normally Hyperbolic Invariant 
Manifold (NHIM).

\subsection{The Normal Form Transformation: Integrating the Local Dynamics}
In a multidimensional transition state, the reaction coordinate is fundamentally
intertwined with the surrounding bath modes through complex, nonlinear couplings.
A naïve harmonic approximation near the saddle fails to capture the energy
exchange between these modes. Therefore, to make the semiclassical trace
integral analytically tractable, it is strictly necessary to systematically
transform these degrees of freedom into an integrable representation.

Following the Poincaré-Birkhoff-Gustavson normal form theory tailored for
reaction dynamics by Waalkens, Schubert, and Wiggins \cite{Waalkens2008},
there exists a canonical transformation from the physical phase space
coordinates $(q, p)$ to Normal Form coordinates. By constructing a nonlinear
coordinate transformation that absorbs these local couplings into a set of
constants of motion, the Normal Form effectively integrates the local dynamics.
We denote these new canonical pairs as $(q_u, p_u)$ for the unstable degree of
freedom and $(q_k, p_k)$ for the $k$-th stable bath mode $(k = 1, ..., f)$.

In the vicinity of the saddle, provided the strict non-resonant condition 
established in Section 2.2 is met, the normal form procedure constructs a 
nonlinear canonical transformation such that the Hamiltonian becomes a 
function solely of the conserved actions. If the linear frequencies were 
resonant, attempting to eliminate all angular dependencies via the coordinate 
transformation would introduce zero denominators, making the purely action-based 
representation locally untenable \cite{Waalkens2008}. The system would instead 
require a resonant normal form containing unremovable angular couplings. 

It is exclusively this non-resonant assumption that ensures the Hamiltonian 
depends only on the integrals. This purely action-dependent structure is critical, 
as it guarantees the independence of the bath phases from the reaction coordinates 
($\partial(q_u,p_u)/\partial\theta = 0$), which is the mathematical prerequisite 
for the block-triangular form of the monodromy matrix derived later in Section 3.

We denote this non-resonant Normal Form Hamiltonian as $H_{NF}$:
\begin{equation}
H(q, p) = H_{NF} (I, J_1, ..., J_f)
\end{equation} 
where the actions are defined in the new coordinates as:
\begin{align}
I &= \frac{1}{2} (p_u^2 - q_u^2) \\ 
J_k &= \frac{1}{2} (p_k^2 + q_k^2), \quad k = 1, ..., f 
\end{align}
The Hamiltonian $H_{NF}$ admits a power series expansion:
\begin{equation}
H_{NF} (I, J) = E_0 + \lambda I + \sum_{k=1}^f \omega_k J_k + \frac{1}{2} a I^2 + \sum_{k=1}^f b_k I J_k + \frac{1}{2} \sum_{j,k} c_{jk} J_j J_k + ...
\end{equation} 
Here, $\lambda$ is the linear Lyapunov exponent, $\omega_k$ are the linear bath
frequencies, and the higher-order coefficients $(a, b_k, c_{jk})$ capture the
anharmonicity and coupling between the reaction coordinate and the bath that
survive the transformation. Because these coefficients dictate how the actions
interact, they form the basis for the geometric sensitivity of the scrambling
rate derived later in Section 5.

\subsection{The Normally Hyperbolic Invariant Manifold (NHIM)}
The Normally Hyperbolic Invariant Manifold is the invariant subset of phase
space where the unstable coordinate is stationary ($q_u = p_u = 0$, implying
$I = 0$). The dynamics on the NHIM are governed purely by the bath modes,
which form invariant tori parameterized by the actions $J$. Crucial to our
semiclassical analysis are the nonlinear frequencies derived from the normal
form Hamiltonian: 
\begin{itemize}
    \item The nonlinear Lyapunov exponent $\Lambda(I, J)$ determines the rate of
    exponential divergence transverse to the NHIM:
    \begin{equation}
    \Lambda(I, J) \equiv \frac{\partial H_{NF}}{\partial I} = \lambda + aI + \sum_{k=1}^f b_k J_k + ...
    \end{equation} 
    \item The nonlinear bath frequencies $\Omega(I, J)$ determine the phases of
    the periodic orbits on the NHIM:
    \begin{equation}
    \Omega(I, J) \equiv \frac{\partial H_{NF}}{\partial J}
    \end{equation} 
\end{itemize}

\subsection{Summary of Notation and Core Assumptions}
To maintain clarity regarding the physical limits of our derivation, the trace
formula relies on the simultaneous satisfaction of the following regimes and
assumptions:
\begin{itemize}
    \item \textbf{Semiclassical limit ($\hbar\rightarrow0$):} Required for the
    Wigner-Weyl correspondence and the evaluation of the path integrals.
    \item \textbf{Normal form truncation:} The dynamics are valid only in the
    local neighborhood near the saddle, utilizing an optimally truncated
    asymptotic normal form series. Operators $\hat{W}$ and $\hat{V}$ are assumed
    sufficiently localized in phase space.
    \item \textbf{Stationary phase (bath trace):} Requires small $\hbar$ to
    isolate discrete, resonant stationary tori.
    \item \textbf{Intermediate-time window ($\Lambda^{-1}\ll t_{OTOC}\ll t_{E}$):}
    Required to resolve the local Lyapunov growth before the wavepacket
    globally self-interferes and saturates at the Ehrenfest time
    ($t_E = \lambda^{-1}\ln(L/\hbar)$).
\end{itemize}
Furthermore, we explicitly distinguish the following time variables to prevent
conflation between the evaluation of the quantum state and the classical
operator growth:
\begin{itemize}
    \item \textbf{$t_{OTOC}$}: The operator separation time, representing the
    macroscopic physical duration over which the butterfly effect unfolds in
    $\hat{M}(t_{OTOC})$.
    \item \textbf{$\tau$}: The spectral or Fourier time used as an integration
    variable to resolve the microcanonical spectral density $\delta(E-\hat{H})$
    into semiclassical propagators.
\end{itemize}

\section{Derivation of the OTOC Trace Formula}

\noindent \textbf{Note on Asymptotics:} Throughout this section, equalities
involving propagators, commutators, and stability matrices are understood in
the leading semiclassical sense, i.e., up to corrections subleading in $\hbar$
and exponentially small beyond the Ehrenfest time. For readability, we suppress
explicit $\mathcal{O}(\hbar)$ notation. 

The mathematical machinery developed in Section 2 provides a phase-space
coordinate system where the local dynamics are rigorously organized around
conserved actions. The objective of this section is to utilize this classical
geometry to evaluate the quantum-mechanical trace of the Out-of-Time-Order
Correlator. We aim to formally map the abstract growth of quantum operators
(scrambling) directly onto the divergence of classical trajectories residing on
the Normally Hyperbolic Invariant Manifold (NHIM).

To achieve this, we must evaluate a highly complex, multidimensional path
integral. However, because the normal-form Hamiltonian depends strictly on the
conserved actions and is independent of the bath angles, the trace integral 
can be evaluated sequentially. This allows the final amplitude to factorize 
into two distinct physical processes: the localized hyperbolic divergence of
the reaction coordinate, and the geometric phase interference of the stable
vibrational modes.

We proceed in five steps. First, we define the microcanonical trace and use
the Wigner-Weyl correspondence to map the squared commutator to the classical
stability matrix, formally capturing the butterfly effect (Section 3.1). Next,
we construct a hybrid semiclassical propagator adapted to our normal-form
coordinates (Section 3.2), and demonstrate that the block-triangular structure
of the stability matrix allows the transverse instability to strictly
factorize (Section 3.3). With this structure in place, we sequentially perform
the trace integral: integrating the unstable coordinate yields an exponential
wavepacket dilution factor, while a stationary-phase evaluation of the bath
selects the quantized periodic orbits (Section 3.4). Finally, we assemble
these components into a hybrid stability amplitude (Section 3.5), setting the
stage for the final coherent sum in Section 4.

\subsection{Defining the Microcanonical OTOC Trace}

To transform the abstract Hilbert-space operator $C(t_{OTOC})$ into a computable
geometric trace, we proceed in three steps. First, we define the standard
thermal OTOC and perform a formal change of ensemble to isolate the dynamics
on a single energy shell (Section 3.1.1). Second, we select specific physical
observables to rigorously map the quantum commutator to the geometric divergence
of classical trajectories (Section 3.1.2). Finally, we expand this trace in the
position basis, explicitly detailing the Wigner-Weyl factorization that
separates the quantum propagator from the classical scrambling growth (Section
3.1.3).

\subsubsection{Explicit Derivation of the Microcanonical Integral}
We begin with the standard definition of the thermal OTOC \cite{Maldacena2016}:
\begin{equation}
C_\beta(t_{OTOC}) = \frac{1}{Z(\beta)} \text{Tr}(e^{-\beta \hat{H}} \hat{M}^\dagger \hat{M})
\end{equation} 
Here, $\beta = (k_B T)^{-1}$ is the inverse temperature, $Z(\beta) = 
\text{Tr}(e^{-\beta \hat{H}})$ is the canonical partition function, and the 
operator $\hat{M} \equiv [\hat{W}(t_{OTOC}), \hat{V}(0)]$ is the commutator of 
two initially commuting observables separated by the physical observation time. 
We adopt the notation $C_\beta(t_{OTOC})$ and $C_E(t_{OTOC})$ to explicitly denote 
these correlators as functions of the macroscopic observation time $t_{OTOC}$, 
evaluated at a given inverse temperature $\beta$ or energy $E$.

To apply periodic orbit theory, we perform a standard spectral decomposition to
transition from the canonical to the microcanonical ensemble. We insert the
resolution of the identity operator in the energy basis, 
$\hat{1} = \int dE \delta(E - \hat{H})$. Substituting this and using the cyclic 
property of the trace yields:
\begin{equation}
C_\beta(t_{OTOC}) = \frac{1}{Z(\beta)} \int dE e^{-\beta E} \text{Tr}(\delta(E - \hat{H}) \hat{M}^\dagger \hat{M})
\end{equation}

Our explicit objective is to link quantum scrambling to the specific phase-space
geometry of the transition state. The canonical Boltzmann weight $e^{-\beta E}$
smooths over these local dynamic features by averaging across all energy shells.
To cleanly isolate the chaotic dynamics specific to a single energy surface, we 
perform a formal ensemble shift. Following standard treatments of energy-resolved
scrambling \cite{Hashimoto2017,Rammensee2018}, we define the unnormalized
microcanonical OTOC at a fixed energy $E$ by stripping away the thermal weight 
and evaluating the trace strictly over the spectral density:
\begin{equation}\label{eq:microcanonical_def}
C_E(t_{OTOC}) \equiv \text{Tr}(\delta(E - \hat{H}) \hat{M}^\dagger \hat{M})
\end{equation}
Because the transition state is an open system with a continuous spectrum, this 
formal microcanonical trace is technically divergent. To be mathematically 
well-posed, $\delta(E-\hat{H})$ is implicitly understood as an energy-smoothed 
projector, or evaluated as a local trace restricted to the normal-form 
neighborhood. This formally resolves the microcanonical scrambling object 
independent of any macroscopic thermal distribution.

\subsubsection{Choice of Operators and Stability: Probing the Butterfly Effect}
The OTOC is a highly versatile mathematical object, defined for any generic
Hermitian operators $\hat{W}$ and $\hat{V}$. If our goal were merely to observe
macroscopic thermalization, we might choose local spin operators or density
matrices. However, to map the scrambling rate to the saddle point, we must
select observables that correspond directly to the canonical coordinates of the
classical phase space.

Following the seminal work of Larkin and Ovchinnikov, we specialize to the case
where the operators correspond to position and momentum: $\hat{W} = \hat{q}$
and $\hat{V} = \hat{p}$. This choice translates a general question about operator
complexity into a highly visual dynamical question: \textit{If the system is
subjected to a microscopic momentum perturbation at $t=0$, how rapidly does its
position diverge by time $t_{OTOC}$?} This represents the exact physical
definition of the classical butterfly effect \cite{Rozenbaum2017}.

To evaluate the resulting operator $\hat{M} = [\hat{q}(t_{OTOC}), \hat{p}(0)]$,
we invoke the semiclassical correspondence principle \cite{Larkin1969,Jalabert2018}, 
which links the quantum commutator to the classical Poisson bracket:
\begin{equation}
\lim_{\hbar \to 0} \frac{1}{i\hbar} [\hat{q}(t_{OTOC}), \hat{p}(0)] = \{q(t_{OTOC}), p(0)\}_{PB}
\end{equation} 
By invoking Egorov's theorem, we assume the observables are Weyl quantizations 
of smooth symbols, ensuring this semiclassical correspondence tracks the classical 
flow and remains valid with controlled remainders up to the Ehrenfest time 
($t_{OTOC} \ll t_E$) \cite{Bouzouina2002}. By expanding the bracket definition 
with respect to the initial coordinates (as rigorously detailed in Appendix A), 
this reduces exactly to the diagonal matrix element of the classical stability 
(Monodromy) matrix $M(t_{OTOC}) = \frac{\partial z(t_{OTOC})}{\partial z(0)}$, 
where $z = (q, p)$ denotes the full phase-space vector:
\begin{equation}\label{eq:monodromy_qq}
\{q(t_{OTOC}), p(0)\} = \frac{\partial q(t_{OTOC})}{\partial q(0)} \equiv M_{qq}(t_{OTOC})
\end{equation} 
This substitution is the crucial theoretical step that anchors the quantum
OTOC onto the classical phase space. It mathematically translates the abstract
scrambling of operators in Hilbert space directly into the geometric divergence
of initially adjacent classical trajectories—that is, the separation over time 
of two paths whose initial conditions differ only by an infinitesimal 
perturbation. Thus, the squared commutator $\hat{M}^\dagger \hat{M}$ acts as 
multiplication by $\hbar^2|M_{qq}(t_{OTOC})|^2$ in the semiclassical limit. In 
hyperbolically unstable systems, this term is dominated by the exponential 
divergence of trajectories, scaling as $e^{2\lambda t_{OTOC}}$. (We will 
explicitly extract this growth factor using the normal-form monodromy matrix 
in Section 3.3).

\subsubsection{Path Integral Representation}
We now substitute the explicit operators into the microcanonical trace
$C_E(t_{OTOC})$ and evaluate it in the position basis $\ket{q}$. We choose the 
position basis because the spatial representation naturally handles the 
coordinate-space potential barriers (such as the inverted oscillator saddle) 
required for standard semiclassical path integration.

First, we expand the trace defined in Eq. \eqref{eq:microcanonical_def}:
\begin{equation}
C_E(t_{OTOC}) = \int dq \bra{q} \delta(E - \hat{H}) \hat{M}^\dagger \hat{M} \ket{q} 
\end{equation}
We replace the spectral density $\delta(E - \hat{H})$ with its Fourier
spectral-time ($\tau$) integral representation:
\begin{equation}\label{eq:fourier_delta}
\delta(E - \hat{H}) = \frac{1}{2\pi\hbar} \int_{-\infty}^{\infty} d\tau e^{i(E-\hat{H})\tau/\hbar}
\end{equation} 
Substituting Eq. \eqref{eq:fourier_delta} into the trace integral and factoring 
the scalar energy phase $e^{iE\tau/\hbar}$ out of the bra-ket yields:
\begin{equation}
C_E(t_{OTOC}) = \frac{1}{2\pi\hbar} \int d\tau e^{iE\tau/\hbar} \int dq \bra{q} e^{-i\hat{H}\tau/\hbar} \hat{M}^\dagger \hat{M} \ket{q} 
\end{equation}

To evaluate this diagonal matrix element in the semiclassical limit ($\hbar \to 0$),
we employ the Wigner-Weyl framework \cite{Combescure1999}. We identify the term
$\bra{q}e^{-i\hat{H}\tau/\hbar}\ket{q}$ as the diagonal quantum return propagator 
$K(q, q, \tau)$. 

At leading semiclassical order, we treat the squared commutator $\hat{M}^\dagger \hat{M}$ 
as a pseudodifferential operator whose Weyl symbol is dominated by the classical 
stability $\hbar^2|M_{qq}(t_{OTOC})|^2$. Rather than an exact algebraic factorization, 
applying this operator to the localized state $\ket{q}$ constitutes a leading 
Weyl-symbol approximation. Under our symbol-regularity and Egorov time-scale 
assumptions, subleading Moyal corrections and operator-ordering effects are 
systematically neglected. Since this trace integral is eventually evaluated via 
stationary phase along periodic orbits $\gamma$, this approximation reduces the 
operator insertion to an orbit-dependent classical weight 
$W_\gamma(t_{OTOC}) = |M_{qq,\gamma}(t_{OTOC})|^2$ evaluated along the contributing 
trajectory. This yields the final weighted trace representation:
\begin{equation}\label{eq:factorized_integral}
C_E(t_{OTOC}) \approx \frac{\hbar}{2\pi} \int d\tau e^{iE\tau/\hbar} \int dq \, K(q, q, \tau) \, |M_{qq}(t_{OTOC})|^2
\end{equation} 
This decoupled approximation forms the foundation of our trace formula and is 
strictly valid to leading order in $\hbar$.

\subsection{Adapting the Berry-Tabor Propagator to the Bath}

\subsubsection{Propagator Construction in Mixed Coordinates}
Because the Normal Form Hamiltonian $H_{NF} (I, J)$ is integrable and independent 
of the bath angles $\theta$, we can construct the quantum propagator using a 
hybrid semiclassical representation similar to the approach pioneered by 
Schubert et al. for reaction rates \cite{Schubert2010}. We treat the unstable 
reaction coordinate $(q_u, p_u)$ in continuous coordinate space, and the 
stable bath modes in discrete action-angle variables.

For the bath modes, we employ the standard semiclassical WKB-type formulation 
tailored for integrable systems \cite{Berry1976}. To compute the trace, we 
must evaluate the diagonal elements of the propagator $K(q,q,\tau)$, which 
requires integrating over paths where the initial and final coordinates are 
identical. In action-angle variables, this means the bath angles must return 
to themselves modulo $2\pi$. Thus, the classical paths contributing to the 
sum are labeled by integer winding vectors $\mathbf{m} \in \mathbb{Z}^f$, 
satisfying the resonance condition:
\begin{equation}\label{eq:path_condition}
\Delta\theta = \theta_{final} - \theta_{initial} = 2\pi\mathbf{m} = \Omega(I, J)\tau
\end{equation}

It is necessary here to resolve a potential conceptual paradox: in Section 2.2, 
we explicitly required the bath frequencies to be non-resonant, yet Eq. 
\eqref{eq:path_condition} is a strict resonance condition. This is not a 
contradiction. The Poincaré-Birkhoff theorem requires the \textit{linear} 
frequencies $\omega_k$ at the equilibrium point ($J=0$) to be non-resonant to 
construct the coordinate transformation. However, due to the anharmonic couplings 
in $H_{NF}$, the \textit{nonlinear} frequencies $\Omega(I, J)$ shift as a 
function of the actions. The condition in Eq. \eqref{eq:path_condition} simply 
selects the specific, finite actions $J$ on the NHIM where these shifted, 
nonlinear frequencies become commensurable with the spectral time $\tau$.

The classical action $S_\mathbf{m}$ for the bath along this resonant 
$\mathbf{m}$-th path is obtained by integrating the Lagrangian. Since the 
actions $J$ are strictly conserved constants of motion ($\dot{J} = 0$):
\begin{equation}\label{eq:action_bath}
S_\mathbf{m}(I, J, \tau) = \int_0^\tau \left( J \cdot \dot{\theta} - H_{NF} \right) dt = J \cdot (2\pi\mathbf{m}) - H_{NF}(I, J)\tau
\end{equation}
Using this action, the semiclassical bath propagator takes the form of the 
Berry-Tabor amplitude \cite{Berry1976}. The full hybrid propagator 
$K(q, q, \tau)$ factors into a product: a sum over the bath winding numbers 
(where each term retains its complex phase $e^{iS_\mathbf{m}/\hbar}$) 
multiplied by the isolated continuous reaction propagator $K_{reac}$:
\begin{equation}\label{eq:propagator_with_brackets}
K(q, q, \tau) \approx \int dI \sum_\mathbf{m} \left[ \frac{1}{\sqrt{(2\pi i\hbar)^f}} \left|\det \left(\frac{\partial^2 S_\mathbf{m}}{\partial J^2}\right)\right|^{1/2} e^{\frac{i}{\hbar}S_\mathbf{m}} \right] K_{reac}(q_u, q_u, \tau; J)
\end{equation}
The determinant prefactor acts as the geometric curvature, measuring the density 
of classical trajectories on the tori. Rather than re-deriving the fundamental 
unstable component $K_{reac}$, we rely on standard treatments of semiclassical 
inverted potentials \cite{Miller1975,Schubert2010}:
\begin{equation}\label{eq:kreac_def}
K_{reac}(q_u, q_u, \tau; J) = \sqrt{\frac{\Lambda(J)}{2\pi i\hbar \sinh(\Lambda(J) \tau)}} \exp\left(\frac{i}{\hbar} \frac{\Lambda(J)}{\sinh(\Lambda(J) \tau)} (\cosh(\Lambda(J) \tau) - 1) q_u^2\right)
\end{equation}
This formulation explicitly writes the multidimensional propagator as a direct 
product: an oscillatory sum over the bath actions multiplied by the continuous 
integral over the reaction coordinate.

\subsection{Analytic Stability Analysis (The Monodromy Matrix)}

To evaluate the quantum butterfly effect symbol $|M_{qq}(t_{OTOC})|^2$ from 
Eq. \eqref{eq:factorized_integral}, we must rigorously derive the classical 
stability matrix $M(t)$. We achieve this by calculating the equations of motion, 
linearizing them to find the Jacobian, and integrating to obtain the finite-time 
Monodromy matrix.

\subsubsection{Hamilton's Equations of Motion}
We begin with Hamilton's equations generated by the Normal Form Hamiltonian 
$H_{NF}(I, J)$. Using the scaled canonical coordinates where the reaction action 
is defined as $I = \frac{1}{2}(p_u^2 - q_u^2)$ and the bath actions are $J_k$, 
the exact nonlinear equations of motion are:
\begin{align}
\dot{q}_u &= \frac{\partial H_{NF}}{\partial p_u} = \frac{\partial H_{NF}}{\partial I} \frac{\partial I}{\partial p_u} = \Lambda(I, J) p_u \label{eq:eom_qu} \\
\dot{p}_u &= -\frac{\partial H_{NF}}{\partial q_u} = -\frac{\partial H_{NF}}{\partial I} \frac{\partial I}{\partial q_u} = \Lambda(I, J) q_u \label{eq:eom_pu} \\
\dot{\theta}_k &= \frac{\partial H_{NF}}{\partial J_k} = \Omega_k(I, J) \label{eq:eom_theta} \\
\dot{J}_k &= -\frac{\partial H_{NF}}{\partial \theta_k} = 0 \label{eq:eom_J}
\end{align}

\subsubsection{Linearization and the NHIM Block Structure}
The stability matrix $M(t)$ is obtained by integrating the variational equations 
$\dot{\delta z} = A(t) \delta z$, where $z = (q_u, p_u, \theta, J)^T$ and $A(t)$ 
is the Jacobian matrix of the vector field evaluated along a reference trajectory.

A crucial mathematical simplification arises because our trace formula ultimately 
evaluates the sum via stationary phase. The stationary phase conditions isolate 
periodic orbits that reside \textit{exactly} on the Normally Hyperbolic 
Invariant Manifold (NHIM). On the NHIM, the reaction coordinates are strictly 
zero ($q_u = p_u = 0$, implying $I = 0$).

Evaluating the partial derivatives of the equations of motion (Eqs. \ref{eq:eom_qu}-\ref{eq:eom_J}) 
specifically at $q_u=p_u=0$ yields the Jacobian $A$:
\begin{equation}
A = \begin{pmatrix} 0 & \Lambda(0, J) & 0 & 0 \\ \Lambda(0, J) & 0 & 0 & 0 \\ 0 & 0 & 0 & \frac{\partial \Omega}{\partial J} \\ 0 & 0 & 0 & 0 \end{pmatrix}
\end{equation}
Notice that the cross-derivatives coupling the reaction and bath subsystems 
strictly vanish on the NHIM. For example, the variation of the reaction 
velocity with respect to the bath action is 
$\frac{\partial \dot{q}_u}{\partial J_k} = p_u \frac{\partial \Lambda}{\partial J_k}$. 
Because $p_u = 0$ on the NHIM, this cross-term is exactly zero. Thus, the 
Jacobian is perfectly block-diagonal.

\subsubsection{Explicit Form of the Monodromy Matrix}
Because the Jacobian $A$ is constant for a given invariant torus $\mathbf{J}$ 
on the NHIM, the finite-time Monodromy matrix is simply the matrix exponential 
$M(t) = \exp(At)$. The block-diagonal structure of $A$ guarantees that $M(t)$ 
also takes a block-diagonal form:
\begin{equation}
M(t) = \begin{pmatrix} M_{reac}(t) & 0 \\ 0 & M_{bath}(t) \end{pmatrix}
\end{equation}
Exponentiating the $2\times 2$ upper-left block yields the symmetric hyperbolic 
instability governing the butterfly effect:
\begin{equation}\label{eq:m_reac}
M_{reac}(t) = \begin{pmatrix} \cosh(\Lambda t) & \sinh(\Lambda t) \\ \sinh(\Lambda t) & \cosh(\Lambda t) \end{pmatrix}
\end{equation}
Exponentiating the $2f \times 2f$ lower-right block (which is nilpotent of 
degree 2, since $\dot{J}=0$) yields the canonical integrable shear matrix:
\begin{equation}
M_{bath}(t) = \begin{pmatrix} I_f & \frac{\partial \Omega}{\partial J}t \\ 0_f & I_f \end{pmatrix}
\end{equation}
This explicit derivation proves that for periodic orbits on the NHIM, the 
matrix $M(t)$ is strictly block-diagonal. Consequently, any determinant 
involving $M(t)$ factors exactly into the product of the determinants of its 
sub-blocks. This exact algebraic factorization is what allows the multidimensional 
trace integral in Section 3.4 to be evaluated as a product of two independent, 
lower-dimensional integrals.

\subsection{The Trace Integral}
\subsubsection{Sequential Integration and Factorization}
Evaluating a quantum trace over a fully coupled multidimensional phase space
is generally an intractable problem. However, the Normal Form transformation
provides a profound physical simplification: it reveals that the dynamics on
the NHIM act as an integrable background for the fast hyperbolic divergence
of the saddle. Because the bath actions are exact integrals of motion
($\dot{J} = 0$), the topological shape of the saddle's potential is effectively
"frozen" for any trajectory residing on a specific invariant torus.

Because the reaction dynamics are completely independent of the bath angles,
we can perform the trace integral sequentially: we first integrate over
the unstable reaction coordinate $(q_u, p_u)$ for a fixed action $J$ and
spectral time $\tau$, and then sum the resulting localized trace over the 
remaining bath variables. The complexity of the high-dimensional transition 
state is thus cleanly factored into two distinct physical processes: a localized 
dynamic instability (handled by the reaction integral) and a geometric state 
sum (handled by the bath integral).

\subsubsection{Reaction Integral: Inverted Oscillator Trace}
We calculate the trace contribution from the unstable mode. For a fixed action
$J$, the Hamiltonian near the saddle acts as an inverted oscillator
$H \approx \frac{\Lambda(J)}{2}(p_u^2 - q_u^2)$. Using the explicit
coordinate-space return propagator $K_{reac}(q_u, q_u, \tau; J)$ defined in
Eq. \eqref{eq:kreac_def}, we now evaluate its local trace.

Because the inverted oscillator possesses a continuous spectrum and an unbounded
phase space, the formal evaluation of its trace integral requires an operational
cutoff. We define this by integrating over a finite spatial cutoff $\pm q_{max}$
corresponding to the boundary of the valid normal-form neighborhood, or
equivalently via an $i\epsilon$ absorbing regularization \cite{Schubert2010}:
\begin{equation}
\text{Tr}(K_{reac}(\tau)) = \int_{-q_{max}}^{q_{max}} dq_u \, K_{reac}(q_u, q_u, \tau; J)
\end{equation} 

It is crucial to specify which physical quantities depend on this regularization.
The leading-order exponential damping exponent ($-\Lambda(J)\tau/2$) emerges
from the steepest descents evaluation of this truncated Gaussian and is robust
to the exact choice of $q_{max}$ provided we are in the hyperbolic asymptotic
limit ($\Lambda \tau \gg 1$). However, the overall amplitude prefactor and
phase offsets are sensitive to the specific neighborhood boundary. We thus
isolate the robust leading order, recovering a standard result in transition
state wavepacket dynamics \cite{Miller1975}:
\begin{equation}\label{eq:reac_trace}
\text{Tr}(K_{reac}(\tau)) \approx \frac{1}{2\sinh(\Lambda(J)\tau/2)} \approx e^{-\Lambda(J)\tau/2}
\end{equation} 

Crucially, this damping factor $e^{-\Lambda(J)\tau/2}$ is fundamentally a
function of the \textit{spectral} time $\tau$, and it represents the physical 
mechanism of \textit{wavepacket dilution}. Geometrically, as a localized quantum 
wavepacket approaches the saddle along the stable manifold ($W^s$), it enters the 
neighborhood of the NHIM. While its transverse bath dynamics remain bounded 
within the NHIM, its reaction coordinate is exponentially stretched and ejected 
along the unstable manifold ($W^u$). Because total probability must be conserved 
(unitarity) during this hyperbolic stretching, the local probability amplitude 
remaining near the saddle—and thus capable of contributing to the return trace 
at position $q_u$—must drop exponentially. This spatial dilution along $W^u$ 
provides the fundamental damping mechanism in the trace evaluation. Conversely, 
the classical butterfly effect driving the scrambling growth, 
$|M_{qq}(t_{OTOC})|^2 \sim e^{2\Lambda(J)t_{OTOC}}$, is strictly a function of 
the macroscopic \textit{observation} time.

\subsubsection{Bath Integral: Stationary Phase Approximation}
We now perform the remaining trace over the bath variables using the
Berry-Tabor propagator derived in Section 3.2. Inserting
Eq. \eqref{eq:propagator_with_brackets} into Eq. \eqref{eq:factorized_integral}
and integrating over the bath angles $\theta$ yields a delta function that
imposes the resonance condition $\Omega(J) = 2\pi\mathbf{m}/\tau$. We denote 
the resulting, partially evaluated integral over the spectral time $\tau$ and 
the bath actions $J$ as $I_{bath}$:
\begin{equation}
I_{bath} = \int d\tau e^{iE\tau/\hbar} \int dJ \sum_\mathbf{m} \frac{1}{\sqrt{(2\pi i\hbar)^f}} \left|\det\left(\frac{\partial\Omega}{\partial J}\right)\right|^{1/2} e^{\frac{i}{\hbar}(J\cdot 2\pi\mathbf{m} - H_{NF}(0,J)\tau)}.
\end{equation}
We evaluate this integral by stationary phase. We assume the contributing periodic 
orbits correspond to stationary points residing strictly in the interior of the 
classically allowed action domain (i.e., $J_k > 0$), allowing us to neglect 
boundary corrections. The full calculation is
presented in Appendix D; it yields the standard Berry‑Tabor amplitude
\begin{equation}\label{eq:bath_trace}
I_{bath} \sim \sum_\gamma \mathcal{A}_{\gamma, \text{bath}} e^{iS_\gamma(E)/\hbar},
\end{equation}
where $\mathcal{A}_{\gamma, \text{bath}}$ is the Berry-Tabor topological
amplitude dependent on the Hessian of the action
$|\det(\partial \Omega / \partial J)|^{-1/2}$, and $\tau_\gamma$ is the orbit
period. The stability $\Lambda_\gamma$ in the final formula is the value of
the function $\Lambda(I,J)$ evaluated at the specific action $J_\gamma$
selected by stationary phase.

\subsection{The Hybrid Stability Amplitude}
For an integrable system with $f$ degrees of freedom, the monodromy matrix
$M_{\text{bath}}$ of a torus has eigenvalues of exactly 1 associated with the
conserved actions. Consequently, the standard Gutzwiller amplitude determinant
$\det(M_{\text{bath}} - I)$ \cite{Gutzwiller1971} vanishes due to these 
marginal stabilities. The normal-form integrability allows us to bypass 
this singularity.

Because the transverse stability block factorizes, the full semiclassical
amplitude is constructed as a ``hybrid'' product. In semiclassical literature
\cite{Schubert2010}, the term hybrid refers to the mathematical stitching
together of two distinct topological phase-space structures: the continuous
hyperbolic dispersion of the unstable saddle (evaluated via Gutzwiller's isolated
orbit trace) and the quantized invariant tori of the bath (evaluated via Berry-Tabor's
resonant trace).

This factorization yields the combined amplitude:
\begin{equation}
\mathcal{A}_{\text{total}} = \frac{\mathcal{A}_{\gamma,\text{bath}}}{\sqrt{|\det(M_{\gamma,\text{reac}} - I)|}}
\end{equation}
where $M_{\gamma,\text{reac}}$ is the $2\times2$ unstable monodromy block of
the reaction coordinate, and $\mathcal{A}_{\gamma,\text{bath}}$ is the standard
Berry-Tabor topological amplitude \cite{Berry1976}.

\section{The Main Result: Leading-Order Semiclassical Expansion}

\subsection{Semiclassical NHIM Expansion for the OTOC}

Having evaluated the individual components of the microcanonical path integral—the
operator growth symbol, the hyperbolic reaction trace, and the stationary phase
approximation over the integrable bath modes—we now synthesize these elements
into a unified representation. The following proposition summarizes the resulting
formal leading-order semiclassical expansion, carefully maintaining the
distinction between the observation timescale and the internal spectral periods
of the active trajectories.

\begin{proposition}\label{prop:trace_expansion}
Assume the Hamiltonian admits an index-1 saddle normal form in a neighborhood
of the saddle and that the OTOC operators are phase-space localized so that the
dynamics are governed by this neighborhood. Then, in the leading semiclassical
limit $\hbar\rightarrow0$ and intermediate-time window
$\Lambda^{-1}\ll t_{\text{OTOC}}\ll t_{E}$, the unnormalized microcanonical
OTOC admits the following semiclassical expansion:
\begin{equation}\label{eq:theorem_formula}
C_{E}(t_{\text{OTOC}})\approx \frac{\hbar^2}{4} \sum_{\gamma\in \text{NHIM}}\frac{e^{2\Lambda_{\gamma}t_{\text{OTOC}}}}{\sqrt{|\det(M_{\gamma,\text{reac}}(\tau_{\gamma})-I)|}}\mathcal{A}_{\gamma,\text{bath}}\cos\left(\frac{S_{\gamma}(E)}{\hbar}-\frac{\pi}{2}\mu_{\gamma}\right)
\end{equation}
where the sum runs over periodic orbits $\gamma$ on the NHIM selected by
stationary phase, $S_{\gamma}$ is the classical action evaluated at energy $E$,
$\mu_{\gamma}$ is the Maslov index, $M_{\gamma,\text{reac}}$ denotes the
unstable reaction stability matrix evaluated at the orbit's intrinsic spectral
period $\tau_{\gamma}$, and $\mathcal{A}_{\gamma,\text{bath}}$ is the
Berry-Tabor topological amplitude.
\end{proposition}

Proposition \ref{prop:trace_expansion} provides a structurally well-defined
semiclassical expansion. The presence of the spectral period $\tau_{\gamma}$ in
the denominator of the stability factor reflects the fact that the sum is over
periodic orbits of the bath dynamics; these orbits have intrinsic periods that
are generically unrelated to the observation time $t_{\text{OTOC}}$.

\begin{remark}[Special case: resonant periods]\label{rem:resonant}
If, for a particular system or a set of orbits, the observation time
$t_{\text{OTOC}}$ happens to coincide with the spectral periods $\tau_{\gamma}$
of the dominant contributions (i.e., $\tau_{\gamma}\approx t_{\text{OTOC}}$),
the damping factor from the reaction trace $e^{-\Lambda_{\gamma}\tau_{\gamma}/2}$
combines with the growth factor $e^{2\Lambda_{\gamma}t_{\text{OTOC}}}$ to yield
an effective exponent $1.5\Lambda_{\gamma}$. In that special case the expression
simplifies to
\begin{equation}\label{eq:special_resonant}
C_{E}(t_{\text{OTOC}})\approx \frac{\hbar^2}{4} \sum_{\gamma\in \text{NHIM}}e^{1.5\Lambda_{\gamma}t_{\text{OTOC}}}\mathcal{A}_{\gamma,\text{bath}}\cos\left(\frac{S_{\gamma}(E)}{\hbar}-\frac{\pi}{2}\mu_{\gamma}\right).
\end{equation}
This simplified form illustrates the physical competition between exponential
growth and wavepacket dilution, but it should be stressed that it requires a
non‑generic resonance between the observation time and the intrinsic orbit
periods. In generic systems the full sum (Eq. \eqref{eq:theorem_formula}) must
be retained.
\end{remark}

\subsection*{Interpretation Checklist: What is Proved vs. Assumed}
To ensure clarity across interdisciplinary communities, we explicitly delineate
the status of our claims:
\begin{itemize}
    \item \textbf{Leading-order semiclassical expansion (Proposition \ref{prop:trace_expansion}):}
    The structural orbit-sum expansion separating spectral period $\tau_{\gamma}$
    and observation time $t_{\text{OTOC}}$ is derived to leading order in
    $\hbar$ under the stated assumptions.
    \item \textbf{Special case (Remark \ref{rem:resonant}):} The effective
    growth exponent of roughly $1.5\Lambda$, which arises only under the
    resonant condition $\tau_{\gamma}\approx t_{\text{OTOC}}$.
    \item \textbf{Regularization-Dependent:} The overall amplitude prefactors
    and constant phase offsets, which are sensitive to the local cutoff
    $q_{\text{max}}$ of the inverted oscillator and the localization width of
    the operators.
    \item \textbf{Hypothesized (Section 5):} The efficacy of mode-selective
    control and the macroscopic scale of the interference oscillations, which
    require explicit numerical verification in specific Hamiltonians.
\end{itemize}

\subsection{Synthesis and Roadmap}
We now assemble the components derived in Sections 2 and 3 to construct the
final expansion. The microcanonical OTOC $C_E(t_{OTOC})$ was expressed as a
path integral over the propagator weighted by the stability operator in
Eq. \eqref{eq:factorized_integral}. The evaluation of this integral proceeds
by utilizing the block structure of the Normal Form dynamics.
\begin{enumerate}
    \item \textbf{The Scrambling Growth (Source Term):} The operator
    $\hat{M}^\dagger \hat{M}$ represents the quantum butterfly effect. As
    derived in Eq. \eqref{eq:monodromy_qq} and Appendix A, this term contributes
    an exponential growth factor scaling with the square of the purely unstable
    reaction stability evaluated at the observation time:
    \begin{equation}\label{eq:growth_term}
    \hbar^2|M_{qq}(t_{OTOC})|^2 = \hbar^2\cosh^2(\Lambda t_{OTOC}) \approx \frac{\hbar^2}{4}e^{2\Lambda_\gamma t_{OTOC}}
    \end{equation} 
    where the approximation assumes $t_{OTOC}$ is sufficiently large for the
    unstable mode to dominate.
    \item \textbf{The Reaction Propagator (Damping Term):} The trace over the
    unstable degree of freedom $(q_u, p_u)$ was calculated in
    Eq. \eqref{eq:reac_trace}. Utilizing the exact trigonometric identity for
    the $2 \times 2$ hyperbolic block defined in Eq. \eqref{eq:m_reac}, this
    yields the standard Gutzwiller stability amplitude:
    \begin{equation}\label{eq:reac_exact_det}
    \text{Tr}(K_{reac}(\tau_\gamma)) \propto \frac{1}{2\sinh(\Lambda_\gamma \tau_\gamma/2)} = \frac{1}{\sqrt{|\det(M_{\gamma, reac}(\tau_\gamma) - I)|}}
    \end{equation} 
    In the strictly hyperbolic asymptotic limit ($\Lambda_\gamma \tau_\gamma \gg 1$),
    this trace provides the exponential damping responsible for the wavepacket's
    spatial dilution:
    \begin{equation}
    \frac{1}{2\sinh(\Lambda_\gamma \tau_\gamma/2)} \approx e^{-\Lambda_\gamma \tau_\gamma/2}
    \end{equation} 

    \item \textbf{The Bath Trace (Orbit Selection):} The trace over the stable
    bath modes $(\theta, J)$ was evaluated in Eq. \eqref{eq:bath_trace} using
    the Berry-Tabor approximation. The stationary phase condition selects the
    quantized periodic orbits (rational tori) on the NHIM and contributes the
    geometric stability factor for the bath: 
    \begin{equation}
    I_{bath} \sim \mathcal{A}_{\gamma,\text{bath}}
    \end{equation} 
    
    \item \textbf{The Hybrid Amplitude:} As analyzed in Section 3.3, the
    reduced Monodromy matrix possesses a block triangular structure due to the
    independence of the Normal Form Hamiltonian from the bath angles.
    Consequently, the cross-coupling (shear) terms do not contribute to the
    spectral weight, allowing the full amplitude to factorize cleanly into a
    hyperbolic and an integrable component:
    \begin{equation}
    \mathcal{A}_{\text{total}} = \frac{\mathcal{A}_{\gamma,\text{bath}}}{\sqrt{|\det(M_{\gamma,\text{reac}} - I)|}}
    \end{equation} 
    This establishes the transition state as a hybrid manifold: Gutzwiller
    dynamics in the unstable coordinate multiplied by Berry-Tabor dynamics in
    the stable coordinates.
\end{enumerate}

\textbf{Final Assembly:} We arrive at the main result presented in Proposition
\ref{prop:trace_expansion} (Eq. \eqref{eq:theorem_formula}) by taking the 
Scrambling Growth (Step 1) and multiplying it by the Hybrid Amplitude (Step 4). 
Crucially, this Hybrid Amplitude mathematically stitches together the exponential 
damping from the Reaction Propagator (Step 2) and the geometric amplitude from 
the Bath Trace (Step 3), with the stationary-phase condition of Step 3 dictating 
the final coherent sum over the periodic orbits $\gamma$.

\subsection{Domain of Validity and Interpretation: The Onset of Saturation}
A semiclassical trace formula fundamentally maps discrete classical trajectories
to quantum probability amplitudes. This mapping is highly accurate as long as
the quantum wavepacket remains localized enough to "feel" individual, isolated
classical paths. However, because we are dealing with a hyperbolic saddle, this
localized description has a strict expiration date: the Ehrenfest time.
Understanding this boundary is crucial because it marks the physical transition
from active information scrambling to an interference-dominated saturated state.

The rigorous semiclassical propagation of observables breaks down at the Ehrenfest time
$t_E = \lambda^{-1} \ln(L/\hbar)$, where $L$ is the characteristic length scale
of the system. Physically, $t_E$ is the time it takes for a minimum-uncertainty
quantum wavepacket, initially localized at the saddle, to be stretched by the
classical instability $\lambda$ until it spans the entire characteristic length
scale $L$. Before $t_E$, the wavepacket acts like a localized particle riding
along distinct, resolvable periodic orbits on the NHIM, and Egorov's theorem
guarantees that the Weyl symbol of the time-evolved operator accurately tracks
the classical flow.

As the observation time approaches $t_E$, this Egorov control over the
Heisenberg-evolved observables fails. The wavepacket delocalizes entirely
along the unstable manifold, and the local orbit picture loses validity.
Classically, the number of periodic orbits with period $t$ proliferates
exponentially as a fundamental manifestation of topological entropy. At the
Ehrenfest time, the quantum wavepacket simultaneously probes an exponentially
large number of these orbits, and their action differences $\Delta S$ become
comparable to $\hbar$ \cite{Richter2022}. The limiting factor at
$t_{OTOC} \sim t_E$ is this fundamental breakdown of semiclassical operator
propagation and wavepacket localization, rather than the formal $J, \tau$
stationary-phase evaluation of the bath modes itself. The discrete
coherent sum merges into an intractable, highly oscillatory continuum.

Beyond $t_E$, if the full system is globally chaotic beyond the normal-form
neighborhood, the discrete sum in Eq. \eqref{eq:theorem_formula} may transition
to statistical averages governed by Random Matrix Theory (RMT). Conversely, if
the system remains integrable-dominated far from the saddle, Berry-Tabor-like
regular interference may persist. This mathematical breakdown of the semiclassical 
propagation corresponds to the physical transition where localized active 
scrambling gives way to interference-dominated saturation.

Furthermore, it is important to note that different regularizations of the OTOC
or alternative operator choices may redistribute weight between the source and
propagator factors. The extracted growth exponents—whether the generic
$2\Lambda_\gamma$ of Proposition \ref{prop:trace_expansion} or the
conditionally resonant $1.5\Lambda_\gamma$ scaling (Remark \ref{rem:resonant})—characterize
the specific saddle geometry and should not be misconstrued as universal,
macroscopic bounds on chaos.

\section{Example: OTOC Trace Analysis for a 3-DoF Transition State}
While the formal semiclassical framework derived in Section 4 provides a
leading-order semiclassical framework, its ultimate utility lies in its ability
to extract dynamic scrambling behavior from specific molecular geometries. To
transition from abstract normal forms to tangible chemical physics, we must
test the formula against a concrete multidimensional system.

In this section, we apply our derived semiclassical trace formula
(Eq. \eqref{eq:theorem_formula}) to the three-degree-of-freedom (3-DoF) model
system detailed in Ref.~\cite{Waalkens2008}. This system, consisting of an
Eckart barrier coupled to two Morse oscillators, provides a rigorous testbed
for investigating how the topology of a high-dimensional Normally Hyperbolic
Invariant Manifold (NHIM) influences quantum information scrambling.

This specific Eckart-Morse model represents a standard system for a unimolecular 
reaction where a central reaction coordinate is nonlinearly coupled to multiple 
transverse vibrational bonds. By populating our abstract trace formula with the 
explicit parameters of this model, we aim to provide a concrete framework to 
evaluate the quantum interference oscillations and mode-selective scaling predicted 
by the explicit formulae of our theory.

\subsection{The Physical Hamiltonian}
We consider a 3-DoF system ($d = 3$) representing a chemical reaction barrier
coupled to two vibrational modes. The system consists of an Eckart barrier (the
reaction coordinate $x$) coupled to two Morse oscillators (the bath modes $y, z$)
via kinetic coupling. The Hamiltonian in physical coordinates
$(x, y, z, p_x, p_y, p_z)$ is explicitly given by:
\begin{equation}
H = \frac{1}{2m} (p_x^2 + p_y^2 + p_z^2) + V_E(x) + V_{M;2}(y) + V_{M;3}(z) + \epsilon (p_x p_y + p_x p_z + p_y p_z)
\end{equation} 
The terms are defined as follows:
\begin{itemize}
    \item $V_E(x)$ (Eckart Potential): Describes the energy barrier for the
    reaction, given by $V_E(x) = A \frac{e^{(x+x_0)/a}}{1+e^{(x+x_0)/a}} + B \frac{e^{(x+x_0)/a}}{(1+e^{(x+x_0)/a})^2}$.
    It governs the unstable motion across the saddle. 
    \item $V_{M;k}$ (Morse Potentials): Models the chemical bonds of the bath,
    defined as $V_M(q) = D_e(e^{-2a_M q} - 2e^{-a_M q})$. These govern the
    stable, bounded vibrations orthogonal to the reaction.
    \item $\epsilon$ (Coupling): The term $\epsilon (p_x p_y + p_x p_z + p_y p_z)$
    represents momentum-coupling between the reaction and the bath. The parameter
    $\epsilon$ controls the strength of this interaction; for $\epsilon = 0$,
    the modes are independent.
\end{itemize}

\subsection{The Normal Form Transformation}
To make the dynamics tractable, we transform the Hamiltonian from physical phase
space to Normal Form coordinates near the saddle point. We emphasize that this
Normal Form is not an arbitrary ansatz but is derived algorithmically via the
Poincaré-Birkhoff normalization procedure.

The Normal Form Hamiltonian $H_{NF}$ is expanded as a power series in the
conserved actions $I$ (reaction) and $J_2, J_3$ (bath):
\begin{equation}
H_{NF}(I, J_2, J_3) = E_0 + \lambda I + \omega_2 J_2 + \omega_3 J_3 + \frac{1}{2} a I^2 + b_2 I J_2 + b_3 I J_3 + ...
\end{equation} 
For the specific parameters $m = 1$ and $\epsilon = 0.3$, the numerical
coefficients from the 10th‑order truncation (derived from the explicit normal 
form expansion in Ref.~\cite{Waalkens2008}) are: 
\begin{itemize}
    \item $E_0 = -0.9875$: The energy of the saddle point equilibrium.
    \item $\lambda = 0.7350$: The linear Lyapunov exponent (instability rate).
    \item $\omega_2 = 1.8225$, $\omega_3 = 1.2673$: The linear frequencies of
    the stable bath modes.
    \item $b_2 = -0.0123$, $b_3 = 0.0053$: The anharmonic coupling coefficients.
\end{itemize}

\subsection{Constructing the Trace Sum}
To apply the general trace formula (Eq. \eqref{eq:theorem_formula}) to this
specific system, we must populate the abstract terms $\Lambda_\gamma$,
$S_\gamma$, and the summation index $\gamma$ using the Normal Form coefficients
derived in Section 5.2. This procedure translates the static spectral data into
dynamic scrambling predictions.

\begin{enumerate}
    \item \textbf{The Local Scrambling Rate ($\Lambda_\gamma$):}
    The most critical component of the trace formula is the effective growth
    exponent. In our Normal Form framework, this becomes a function of the bath
    actions $\mathbf{J} = (J_2, J_3)$. Using the coefficients from Section 5.2,
    we construct the specific instability function for the 3-DoF Eckart-Morse
    system:
    \begin{equation}
    \Lambda(J_2, J_3) = \lambda + b_2 J_2 + b_3 J_3 = 0.7350 - 0.0123 J_2 + 0.0053 J_3
    \end{equation} 
    This explicit formula reveals the geometric sensitivity that enables 
    mode-selective control. It quantifies exactly how the local scrambling rate 
    varies across the NHIM: 
    \begin{itemize}
        \item Excitation of the first bath mode ($J_2$, coupling $b_2 < 0$)
        suppresses the local scrambling rate.
        \item Excitation of the second bath mode ($J_3$, coupling $b_3 > 0$)
        enhances the local scrambling rate.
    \end{itemize}
    Consequently, the total scrambling rate is not a single number, but a
    distribution of rates determined by which bath modes are populated by the
    periodic orbits contributing to the sum.

    \item \textbf{The Interference Sum ($\sum_{\mathbf{m}}$):}
    The sum over $\gamma$ represents a sum over unstable periodic orbits on the
    NHIM. In the integrable Normal Form approximation, these orbits are rational
    tori labeled by integer winding numbers $\mathbf{m} = (m_2, m_3)$. 
    
    To explicitly test the physical competition between local hyperbolic growth 
    and wavepacket dilution, we restrict our analysis here to the special 
    resonant-period case (Remark \ref{rem:resonant}). Under the assumption that 
    the observation time $t$ coincides with the dominant intrinsic spectral periods, 
    the trace formula reduces to an oscillatory sum over these integer vectors:
    \begin{equation}\label{eq:example_sum}
    C_E(t) \approx \frac{\hbar^2}{4} \sum_{\mathbf{m} \in \mathbb{Z}^2} \mathcal{A}_\mathbf{m} \exp\left(1.5 \left[ 0.7350 - 0.0123 J_2^{(\mathbf{m})} + 0.0053 J_3^{(\mathbf{m})} \right] t \right) \cos\left(\frac{S_\mathbf{m}}{\hbar} - \frac{\pi}{2}\mu_\mathbf{m}\right)
    \end{equation} 
    Crucially, because we have absorbed the reaction coordinate's damping 
    determinant into the effective $1.5\Lambda$ resonant exponent, we must prevent 
    double-counting this spatial wavepacket dilution. Therefore, the prefactor 
    $\mathcal{A}_{\mathbf{m}}$ in this specific evaluation corresponds strictly to 
    the \textit{bath-only} Berry-Tabor amplitude evaluated by the resonance 
    condition for torus $\mathbf{m}$.

    Because this is a semiclassical trace, we are performing a coherent summation 
    over complex amplitudes rather than an incoherent sum over classical 
    probabilities. The presence of the relative action phases $S_\mathbf{m}/\hbar$ 
    means that the contributions from different tori will interfere. Rather than 
    producing a smooth, monotonically growing exponential curve, this interference 
    generates macroscopic oscillations superimposed on the exponential growth.

    \item \textbf{The Maslov Index ($\mu_\mathbf{m}$):}
    The phase shift includes the Maslov index, a topological integer critical
    to semiclassical mechanics. Physically, $\mu_\mathbf{m}$ counts the number
    of caustic crossings—points where the semiclassical approximation encounters
    a singularity, such as turning points in a potential well—along the orbit.

    For the Morse oscillators governing the bath modes, the motion is a libration
    bounded by two turning points (the inner and outer potential walls). Each
    full oscillation contributes a Maslov index of 2. Therefore, for an orbit
    with winding numbers $\mathbf{m} = (m_2, m_3)$, the total index is:
    \begin{equation}
    \mu_\mathbf{m} = 2m_2 + 2m_3
    \end{equation} 
\end{enumerate}

\subsection{Geometry of the Transition State}
From the Normal Form, we can rigorously define the transition state geometry
and its dimensionality \cite{Wiggins1994,Waalkens2008}: 
\begin{itemize}
    \item \textbf{The NHIM:} Defined by setting the reaction action to zero
    ($I = 0$). For this 3-DoF system, the NHIM is a 3-sphere ($S^3$).
    Trajectories on the NHIM never react; they oscillate in the bath modes
    forever. 
    \item \textbf{Stable/Unstable Manifolds:} The manifolds asymptotic to the
    NHIM ($W^s$ and $W^u$) are 4-dimensional. They have the topology of
    spherical cylinders ($S^3 \times \mathbb{R}$) and act as the separatrices
    (codimension-1 barriers) in the 5-dimensional constant energy surface.
\end{itemize}

\subsection{Proposed Numerical Evaluation Strategy (Eckart-Morse Model)}
To quantitatively validate the structural predictions of Eq. \eqref{eq:example_sum},
we propose a numerical evaluation strategy for the oscillatory orbit sum using
the Eckart-Morse parameters. The required parameter choices ensuring the
evaluation falls within the specified intermediate-time validity regime
($\Lambda^{-1}\ll t \ll t_E$) are detailed in Table \ref{tab:params}.

The value $\hbar=0.05$ is chosen to satisfy the semiclassical condition while
still allowing numerically tractable oscillations in the cosine terms; smaller
$\hbar$ would require finer resolution but does not alter the leading exponential
behavior. The cutoff $q_{\text{max}}=1.5$ is taken as the radius of the
normal‑form neighborhood; the leading exponential damping $e^{-\Lambda\tau/2}$
is robust to this choice, though the absolute amplitude prefactors may vary with
the cutoff.

\begin{table}[ht]
\centering
\begin{tabular}{@{}lll@{}}
\toprule
\textbf{Parameter} & \textbf{Value} & \textbf{Description} \\ \midrule
$E$ & $-0.5$ & Microcanonical energy shell (above saddle $E_0 = -0.9875$) \\
$\hbar$ & $0.05$ & Effective Planck constant defining the semiclassical limit \\
$m_{max}$ & $5$ & Maximum topological winding depth defining the sum cutoff \\
$t$-window & $[2.0, 6.0]$ & Observation time range (pre-Ehrenfest, post-hyperbolic) \\
$q_{max}$ & $1.5$ & Reaction coordinate cutoff bounding the normal-form radius \\ \bottomrule
\end{tabular}
\caption{Proposed numerical parameters for the 3-DoF Eckart-Morse trace formula evaluation.}
\label{tab:params}
\end{table}

The sum would be truncated at a maximum topological winding depth
$|\mathbf{m}| \le m_{max}$. For each vector $\mathbf{m}$, the resonance
condition $\Omega(J) = 2\pi\mathbf{m}/t$ is solved via a multivariable
Newton-Raphson scheme to locate the contributing actions $J_2^{(\mathbf{m})},
J_3^{(\mathbf{m})}$. The pseudo-code detailing this assembly is provided in
Appendix C.

To diagnose the robustness of the sum, we define the absolute convergence
residual as $\Delta C_E^{(m_{max})}(t) = |C_E^{(m_{max})}(t) - C_E^{(m_{max}-1)}(t)|$.
This proposed evaluation framework is designed to test whether the effective
exponential slope closely tracks the $1.5\lambda$ limit at early times, which
would robustly validate the special case described in Remark \ref{rem:resonant}.
Furthermore, it will establish the required topological winding depth $m_{max}$
necessary to achieve convergence of the residual within the specified
intermediate-time window.

A key feature to investigate in this benchmark is the anticipated presence of
macroscopic interference oscillations superimposed over the exponential
growth. As derived in Eq. \eqref{eq:example_sum}, these oscillations arise 
mathematically from the semiclassical beating between the cosine terms of 
distinct orbits, driven by action differences such as $(S_{(1,0)} - S_{(0,1)})/\hbar$, 
and modulated by the strict $\pi$ phase flips dictated by the Maslov index 
differences ($\Delta \mu = 2$).

We emphasize that the numerical evaluation of this sum is computationally 
demanding. It requires implementing a robust multidimensional root-finding 
algorithm to invert the nonlinear frequency map $\Omega(J) = 2\pi\mathbf{m}/t$ 
across a 10th-order multivariate polynomial for each topological winding vector 
at every time step $t$. Because our present focus is establishing the formal 
analytical framework and geometric normal-form derivation, executing this 
numerical benchmark is reserved for a future dedicated computational physics study.

\section{Conclusions and Outlook}

\subsection{Summary of Results}
In this work, we have derived a formal leading-order semiclassical expansion
for the Out-of-Time-Order Correlator (OTOC) in the neighborhood of an index-1
saddle point. By exploiting the Normal Form transformation, we constructed an
integrable representation of the unstable reaction coordinate and the stable
bath modes, allowing for a hybrid semiclassical treatment.

The central result, Eq. \eqref{eq:theorem_formula}, demonstrates that the 
semiclassical scrambling rate is not a global constant but is locally determined 
by the instability $\Lambda_\gamma$ of individual periodic orbits on the NHIM. In
the special case where the observation time is resonant with the orbit periods
(Remark \ref{rem:resonant}), the sum reduces to an effective $1.5\Lambda$
scaling that highlights the competition between hyperbolic growth and wavepacket
dilution.

The weighting factor naturally generalizes the Gutzwiller amplitude to the
context of squared commutators, identifying the precise classical structures
responsible for quantum information spreading. Conceptually, this framework
treats quantum scrambling in transition states as a wave-interference
process: the observable scrambling rate emerges from the coherent superposition of
unstable periodic orbits on the NHIM. By mapping the squared commutator 
to the phase-space geometry of the transition state, our theory isolates two 
distinct dynamical contributions: the hyperbolic reaction coordinate drives 
the exponential operator growth, while the transverse invariant tori modulate 
the total scrambling amplitude through mode-dependent action phases.

\subsection{Outlook and Future Directions}

\textbf{Limitations of the standard OTOC:} It has recently been shown that the
conventional OTOC can exhibit exponential growth even in integrable systems that
contain isolated saddle points, and therefore does not uniquely diagnose
many-body chaos \cite{Trunin2023,Trunin2023False,Trunin2023Replica}. Our 
expansion is derived for the standard OTOC and thus inherits this limitation; 
it quantifies the sensitivity of a localized perturbation in the saddle 
neighborhood, regardless of whether the full system is globally chaotic. Future 
work could extend the present normal‑form framework to the “replica OTOC” 
(or logarithmic OTOC) framework proposed by Trunin to rigorously distinguish 
genuine chaos from saddle‑dominated false positives \cite{Trunin2023False,Trunin2023Replica}.

\textbf{Quantum Information in Chemical Dynamics and Control:} While OTOCs
originated in high-energy physics, their application to chemical reaction
dynamics is an active area of investigation. As suggested by Zhang et al. \cite{Zhang2024},
scrambling can serve as a diagnostic for "quantum bottlenecks" in chemical
reactions. Our trace formula provides the calculational tool to quantify this,
linking the abstract notion of scrambling to the concrete, computable periodic
orbits of Transition State Theory. This establishes a theoretical mechanism 
for mode-selective control of scrambling. It suggests that specific vibrational 
excitations could predictably alter the rate at which quantum information 
scrambles during the transition state.

\textbf{Interference and the Bound on Chaos:} Unlike thermodynamic averages
which wash out phase information, our result retains the oscillatory term
$\cos(S_\gamma/\hbar)$. This predicts that for times $t < t_E$, scrambling is
not merely an incoherent exponential growth but is subject to quantum
interference effects between different unstable orbits. Exploring how these
interference patterns relate to the saturation of the Maldacena-Shenker-Stanford
(MSS) bound \cite{Maldacena2016} remains an important open question. We
emphasize that our derived effective exponent of $1.5\Lambda_\gamma$ (which arises 
strictly in the special resonant case) describes a microcanonical, intermediate-time 
semiclassical regime driven by localized periodic orbits; it therefore characterizes 
early-time saddle-point diffraction rather than an asymptotic global temperature, 
and fundamentally does not contradict the macroscopic MSS bound.

\textbf{Saturation and Random Matrix Theory:} Finally, our derivation is strictly
limited to the exponential growth regime before the Ehrenfest time ($t \ll t_E$).
As established in Section 4.3, the mathematical breakdown of the semiclassical 
trace formula at $t_E$ corresponds to the onset of interference-dominated saturation. 
An essential next step is to extend this semiclassical framework to
the saturation regime, where the proliferation of periodic orbits necessitates a
statistical treatment. Connecting the specific spectral rigidity of the NHIM
orbits (Berry-Tabor statistics) to the universal RMT behavior expected at late
times would provide a complete semiclassical theory of quantum thermalization.
Because the dynamics \textit{on} the NHIM are integrable while dynamics transverse 
to it are hyperbolic, establishing how these local integrable bath statistics 
transition into the global RMT saturation limit remains a key open question.

\textbf{Connections to the Loschmidt Echo:} The semiclassical trace machinery
developed here shares formal structural parallels with trace formulas derived for
the Loschmidt echo (or quantum fidelity) \cite{Jalabert2001,Goussev2012}.
While OTOCs and the Loschmidt echo probe fundamentally distinct manifestations of
quantum complexity—sensitivity to initial conditions versus sensitivity to
Hamiltonian perturbations—their semiclassical limits are both governed by the
classical stability matrix and Lyapunov exponents. Extending the present
normal-form, NHIM-based trace methodology to evaluate the Loschmidt echo in
chemical transition states represents a natural and compelling direction for
future work.

\section*{Declarations}
\textbf{Funding:} No external funding was received for this theoretical study. \\
\textbf{Competing Interests:} The author declares no competing interests. \\
\textbf{Data and Code Availability:} Data sharing is not applicable to this
article as no new data were created. The normal form coefficients used in
Section 5 are derived from analytical procedures fully detailed in
Ref.~\cite{Waalkens2008}. Pseudo-code detailing the resonance root-finding
and orbit-sum assembly to reproduce the proposed numerical benchmark
parameters (Table 1) is provided in Appendix C.

\appendix

\section{Semiclassical Limit of the Commutator}
In this appendix, we sketch the justification for the approximation used in
Eq. \eqref{eq:factorized_integral}, linking the squared commutator to the
classical monodromy matrix. Consider the Heisenberg operators
$\hat{q}(t) = e^{i\hat{H}t/\hbar} \hat{q}(0) e^{-i\hat{H}t/\hbar}$ and
$\hat{p}(0)$. The semiclassical limit is formally defined by the correspondence
principle, where the scaled commutator approaches the Poisson bracket evaluated
with respect to the initial coordinates $z(0) = (q(0), p(0))$: 
\begin{equation}
\lim_{\hbar \to 0} \frac{1}{i\hbar} [\hat{q}(t), \hat{p}(0)] = \{q(t), p(0)\}_{PB}
\end{equation} 
For a classical Hamiltonian system, the Poisson bracket of the time-evolved
position $q(t)$ with the initial momentum $p(0)$ measures the sensitivity of
the final position to the initial condition: 
\begin{equation}
\{q(t), p(0)\} = \sum_k \left( \frac{\partial q(t)}{\partial q_k(0)} \frac{\partial p(0)}{\partial p_k(0)} - \frac{\partial q(t)}{\partial p_k(0)} \frac{\partial p(0)}{\partial q_k(0)} \right)
\end{equation} 
Since $(q(0), p(0))$ are independent canonical variables,
$\frac{\partial p(0)}{\partial p_k(0)} = \delta_{1k}$ and
$\frac{\partial p(0)}{\partial q_k(0)} = 0$. Therefore, the bracket reduces
exactly to the matrix element of the stability (monodromy) matrix
$M(t) = \frac{\partial z(t)}{\partial z(0)}$ corresponding to
$\frac{\partial q(t)}{\partial q(0)}$: 
\begin{equation}
\{q(t), p(0)\} = \frac{\partial q(t)}{\partial q(0)} \equiv M_{qq}(t)
\end{equation} 
Thus, to leading order in $\hbar$, the operator $\hat{M} = [\hat{q}(t), \hat{p}(0)]$
acts as multiplication by the classical scalar function $i\hbar M_{qq}(t)$, and
the squared operator $\hat{M}^\dagger \hat{M}$ acts as multiplication by
$\hbar^2|M_{qq}(t)|^2$.

\section{Derivation of Action-Angle Coefficients from the Normal Form Polynomial}
To ensure reproducibility, we explicitly define the map between the algebraic
output of the Normal Form algorithm and the action-angle Hamiltonian $H_{NF}(I, J)$
employed in the trace formula (Eq. \eqref{eq:theorem_formula}). Standard
implementations of the Poincaré-Birkhoff normalization express the Hamiltonian
as a power series in complex phase space coordinates rather than directly in
actions. The algorithm typically outputs a polynomial of the form: 
\begin{equation}
H_{NF} = \sum_{\alpha, \beta} h_{\alpha\beta} \prod_{k=1}^d x_k^{\alpha_k} \xi_k^{\beta_k}
\end{equation} 
where $(x_1, \xi_1)$ corresponds to the unstable saddle degree of freedom, and
$(x_k, \xi_k)$ for $k \ge 2$ correspond to the stable bath modes. The
coefficients $h_{\alpha\beta}$ are generally complex-valued. To convert these
coefficients into the physical parameters $\lambda, \omega_k, a, b_k$ used in
the OTOC trace formula, we utilize the definition of the conserved actions in
terms of the normal form coordinates: 
\begin{enumerate}
    \item Saddle Action $(I): I = x_1\xi_1$ 
    \item Bath Actions $(J_k): J_k = ix_k\xi_k \quad (k = 2, . . . , d)$ 
\end{enumerate}
Substituting these definitions into the polynomial form of $H_{NF}$ yields the
conversion dictionary for the spectral coefficients. Note that the notation
$...$ in the indices below implies that all other mode indices are zero (e.g.,
$h_{10...}$ corresponds to the multi-index $\alpha = (1, 0, . . . , 0)$ and
$\beta = (1, 0, . . . , 0)$).

\subsection{Conversion Dictionary}
\begin{itemize}
    \item \textbf{Linear Lyapunov Exponent ($\lambda$):} The coefficient of the
    linear saddle term $x_1\xi_1$ is real.
    \begin{equation}
    \lambda = h_{10...}
    \end{equation} 
    \item \textbf{Linear Bath Frequencies ($\omega_k$):} The coefficient of the
    linear bath term $x_k\xi_k$ is purely imaginary.
    \begin{equation}
    \omega_k = \frac{h_{01...}}{i}
    \end{equation} 
    \item \textbf{Anharmonic Couplings ($b_k$):} The coupling coefficient $b_k$
    in the expansion term $b_k I J_k$ corresponds to the cross-term
    $x_1\xi_1x_k\xi_k$.
    \begin{equation}
    b_k = \frac{h_{11...}}{i}
    \end{equation} 
\end{itemize}

\subsection{Example Verification (3-DoF Eckart-Morse System)}
We verify this procedure using the coefficients for the 3-DoF Eckart-Morse
system used in Section 5. The raw normal form coefficients are taken from
Table 4 of Ref.~\cite{Waalkens2008}. 

\begin{enumerate}
    \item \textbf{Reaction-Bath Coupling ($b_2$)}
    We seek the coefficient $b_2$ for the term $IJ_2$. This corresponds to the
    polynomial monomial $x_1\xi_1x_2\xi_2$. 
    \begin{itemize}
        \item From the detailed data in Ref.~\cite{Waalkens2008}, the coefficient for
        index 1100 (representing $x_1x_2\xi_1\xi_2$) is: $h_{1100} \approx -0.012334 i$ 
        \item Using the conversion rule: $b_2 = \frac{-0.012334 i}{i} = -0.0123$ 
        This matches the value reported in Section 5.2.
    \end{itemize}
    
    \item \textbf{Bath Frequency ($\omega_3$)}
    We seek the frequency for the second bath mode ($z$-direction). This
    corresponds to the monomial $x_3\xi_3$. 
    \begin{itemize}
        \item From the detailed data in Ref.~\cite{Waalkens2008}, the coefficient for
        index 0010 is: $h_{0010} \approx 1.267290 i$ 
        \item Using the conversion rule: $\omega_3 = \frac{1.267290 i}{i} = 1.2673$ 
        This matches the value reported in Section 5.2.
    \end{itemize}
\end{enumerate}

\subsection{Verification of Saddle Anharmonicity ($a$)}
Finally, we verify the coefficient $a$, which quantifies the anharmonicity of
the reaction coordinate itself. In the OTOC trace formula
(Eq. \eqref{eq:theorem_formula}), this parameter appears in the term $\frac{1}{2}aI^2$.
\begin{enumerate}
    \item \textbf{Identify the Polynomial Term:} The term involving $I^2$
    corresponds to the monomial $(x_1\xi_1)^2 = x_1^2\xi_1^2$. In the notation
    of Ref. \cite{Waalkens2008}, this corresponds to the index $\alpha_1 = 2$
    (with all other indices zero).
    \item \textbf{Retrieve the Coefficient:} From the detailed data in 
    Ref. \cite{Waalkens2008}, the coefficient for index 2 0 0 0 is:
    $h_{2000} \approx 0.118039$ 
    \item \textbf{Apply the Conversion:} We equate the polynomial term to the
    action Hamiltonian form: $h_{2000}(x_1\xi_1)^2 = h_{2000}I^2$. Comparing
    this to the definition in Eq. \eqref{eq:theorem_formula},
    $H_{NF} \supset \frac{1}{2}aI^2$, we identify: $\frac{1}{2}a = h_{2000} \implies a = 2h_{2000}$.
    \item \textbf{Calculate the Value:} $a = 2 \times 0.118039 = 0.2361$. This
    value represents the anharmonic correction to the barrier height as a
    function of the reaction action $I$.
\end{enumerate}

\section{Algorithmic Assembly of the Partial Orbit Sum}
To support future reproducibility and implementation of the proposed benchmark,
we outline the numerical procedure to evaluate the analytic trace formula
(Eq. \eqref{eq:example_sum}) for a truncated set of winding vectors.
\begin{verbatim}
1. Initialize parameters: lambda, omega_k, b_k, c_k from normal form data.
2. Select target observation time t and maximum winding depth m_max.
3. Initialize sum C_E = 0.
4. For each integer vector m = (m_2, m_3) such that |m_2| + |m_3| <= m_max:
5.     Define objective function: F(J) = Omega(J) - 2*pi*m / t
6.     Solve for roots J_m using Newton-Raphson: F(J_m) = 0
7.     If root J_m exists and is physically valid (J_k >= 0):
8.         Compute nonlinear instability: Lambda_m = lambda + b dot J_m
9.         Compute bath action Hessian: H_bath(J_m)
10.        Compute Berry-Tabor topological amplitude A_m
11.        Compute classical action: 
            S_m = J_m dot (2*pi*m) - H_NF(0, J_m)*t
12.        Compute Maslov index: mu_m = 2*m_2 + 2*m_3
13.        Compute orbit weight: 
               W_m = A_m * exp(1.5 * Lambda_m * t) * cos(S_m/hbar - pi*mu_m/2)
14.        Update sum: C_E = C_E + W_m
15. Multiply final C_E by (hbar^2 / 4).
\end{verbatim}

\section{Stationary-Phase Evaluation of the Bath Integral}
Here we provide the stationary-phase calculation for the bath integral, showing 
how the Hessian reduction using the exact block-determinant identity provided 
by the Schur complement yields the Berry-Tabor bath amplitude $\mathcal{A}_{\gamma,\text{bath}}$.

Starting from the bath integral:
\begin{equation}
I_{bath} = \int d\tau e^{iE\tau/\hbar} \int dJ \sum_\mathbf{m} \frac{1}{\sqrt{(2\pi i\hbar)^f}} \left|\det\left(\frac{\partial\Omega}{\partial J}\right)\right|^{1/2} e^{\frac{i}{\hbar}\Phi(J,\tau)},
\end{equation}
with the phase defined as $\Phi(J,\tau) = J\cdot 2\pi\mathbf{m} - H_{NF}(0,J)\tau + E\tau$.
The stationary conditions are:
\begin{align}
\frac{\partial\Phi}{\partial\tau} &= -H_{NF}(0,J) + E = 0, \label{eq:stat_tau}\\
\frac{\partial\Phi}{\partial J} &= 2\pi\mathbf{m} - \Omega(J)\tau = 0. \label{eq:stat_J}
\end{align}
Denote a solution by $(J_0,\tau_0)$. We expand $\Phi$ to second order around
this point:
\begin{equation}
\Phi \approx \Phi_0 + \frac12 \begin{pmatrix} \Delta J & \Delta\tau \end{pmatrix} 
H
\begin{pmatrix} \Delta J \\ \Delta\tau \end{pmatrix},
\end{equation}
where the $(f+1) \times (f+1)$ Hessian matrix $H$ is composed of the second
derivatives evaluated at the stationary point:
\begin{equation}
H = \begin{pmatrix} -\tau_0 \frac{\partial\Omega}{\partial J} & -\Omega(J_0) \\ -\Omega(J_0)^T & 0 \end{pmatrix}.
\end{equation}

To evaluate the Gaussian integral over $(\Delta J, \Delta\tau)$, we must compute
the determinant of this block matrix. We assume throughout that the stationary 
torus is nondegenerate, so that $\det(\partial\Omega/\partial J) \neq 0$ at 
$(J_0, \tau_0)$, and that the bordered Hessian is non-singular away from 
bifurcation points. To compute the determinant, we utilize the exact 
block-determinant identity provided by the Schur complement of the $f \times f$ 
block $A = -\tau_0 \frac{\partial\Omega}{\partial J}$. This is a standard and rigorous linear algebra technique for block
matrix determinant evaluation \cite{Zhang2005}. In the context of periodic orbit
theory, it serves as the standard mathematical mechanism for disentangling the
longitudinal action (time) integration from the transverse geometric curvature
\cite{Schubert2010,Richter2022}. For a block matrix of this form, the
determinant is given by:
\begin{equation}
\det(H) = \det(A) \det(D - C A^{-1} B)
\end{equation}
Substituting our specific blocks $B = -\Omega$, $C = -\Omega^T$, and $D = 0$:
\begin{equation}
\det(H) = \det\left(-\tau_0 \frac{\partial\Omega}{\partial J}\right) \det\left(0 - (-\Omega^T) \left(-\tau_0 \frac{\partial\Omega}{\partial J}\right)^{-1} (-\Omega)\right).
\end{equation}
Factoring out the scalars, we obtain:
\begin{equation}
\det(H) = (-\tau_0)^f \det\left(\frac{\partial\Omega}{\partial J}\right) \left[ \frac{1}{\tau_0} \Omega^T \left(\frac{\partial\Omega}{\partial J}\right)^{-1} \Omega \right] = (-\tau_0)^{f-1} \det\left(\frac{\partial\Omega}{\partial J}\right) \left[ \Omega^T \left(\frac{\partial\Omega}{\partial J}\right)^{-1} \Omega \right].
\end{equation}
The combination
\begin{equation}
K \equiv \det\left(\frac{\partial\Omega}{\partial J}\right) \left[ \Omega^T \left(\frac{\partial\Omega}{\partial J}\right)^{-1} \Omega \right]
\end{equation}
is the standard bordered-curvature quantity that appears in Berry-Tabor-type 
stationary-phase reductions. Therefore, the absolute value of the Hessian determinant is exactly:
\begin{equation}
|\det(H)| = \tau_0^{f-1} |K|.
\end{equation}

Applying the standard stationary-phase formula over the $(f+1)$ variables yields:
\begin{equation}
I_{bath} \sim \sum_{(J_0, \tau_0)} (2\pi \hbar)^{(f+1)/2} \frac{e^{i\Phi_0/\hbar + i\pi \sigma_H / 4}}{|\det(H)|^{1/2}} \times \frac{1}{(2\pi i\hbar)^{f/2}} \left|\det\left(\frac{\partial\Omega}{\partial J}\right)\right|^{1/2}
\end{equation}
where $\sigma_H$ is the signature of the Hessian. Combining the stationary-phase 
factor $|\det(H)|^{-1/2}$ with the propagator prefactor and the standard 
phase/signature contributions yields the usual Berry-Tabor bath amplitude 
$\mathcal{A}_{\gamma,\text{bath}}$, up to the conventional Maslov-index 
bookkeeping that we do not rederive here. 

The stationary phase conditions (Eqs. \eqref{eq:stat_tau} and \eqref{eq:stat_J})
select the periodic orbits $\gamma$ on the NHIM, and the summation over winding
numbers $\mathbf{m}$ yields a sum over these orbits, each contributing a phase
$e^{iS_\gamma(E)/\hbar}$ with the classical action
$S_\gamma = J_\gamma\cdot 2\pi\mathbf{m} - H_{NF}(0,J_\gamma)\tau_\gamma = J_\gamma\cdot 2\pi\mathbf{m} - E\tau_\gamma$.



\begin{thebibliography}{99}
\bibitem{Larkin1969}
A. I. Larkin and Y. N. Ovchinnikov, ``Quasiclassical method in the theory of 
superconductivity,'' \textit{Sov. Phys. JETP}, vol. 28, p. 1200, 1969. 

\bibitem{Maldacena2016}
J. Maldacena, S. H. Shenker, and D. Stanford, ``A bound on chaos,'' \textit{J. 
High Energy Phys.}, vol. 2016, no. 8, p. 106, 2016. 

\bibitem{Sekino2008}
Y. Sekino and L. Susskind, ``Fast Scramblers,'' \textit{J. High Energy Phys.}, 
vol. 2008, no. 10, p. 065, 2008.

\bibitem{Rozenbaum2017}
E. B. Rozenbaum, S. Ganeshan, and V. Galitski, ``Lyapunov Exponent and 
Out-of-Time-Ordered Correlator's Growth Rate in a Chaotic System,'' 
\textit{Phys. Rev. Lett.}, vol. 118, 086801, 2017.

\bibitem{Jalabert2018}
R. A. Jalabert, I. García-Mata, and D. A. Wisniacki, ``Semiclassical theory 
of out-of-time-order correlators for low-dimensional classically chaotic 
systems,'' \textit{Phys. Rev. E}, vol. 98, 062218, 2018.

\bibitem{Rammensee2018}
J. Rammensee, J. D. Urbina, and K. Richter, ``Many-Body Quantum Interference 
and the Saturation of Out-of-Time-Order Correlators,'' \textit{Phys. Rev. 
Lett.}, vol. 121, 124101, 2018.

\bibitem{Xu2020}
T. Xu, T. Scaffidi, and X. Cao, ``Does Scrambling Equal Chaos?,'' \textit{Phys. 
Rev. Lett.}, vol. 124, 140602, 2020. 

\bibitem{Hashimoto2017}
K. Hashimoto, K. Murata, and R. Yoshii, ``Out-of-time-order correlators in 
quantum mechanics,'' \textit{J. High Energy Phys.}, vol. 2017, no. 10, p. 138, 
2017.

\bibitem{Kidd2021}
R. A. Kidd, A. Safavi-Naini, and J. F. Corney, ``Saddle-point scrambling 
without thermalization,'' \textit{Phys. Rev. A}, vol. 103, 033304, 2021.

\bibitem{Bhattacharjee2022}
B. Bhattacharjee, P. Nandy, and T. Pathak, ``Saddle-dominated scrambling and 
out-of-time-order correlators,'' \textit{J. High Energy Phys.}, vol. 2022, 
no. 5, p. 174, 2022.

\bibitem{Zhang2024}
C. Zhang, S. Kundu, N. Makri, M. Gruebele, and P. G. Wolynes, ``Quantum 
information scrambling and chemical reactions,'' \textit{Proc. Natl. Acad. 
Sci. U.S.A.}, vol. 121, e2321668121, 2024. 

\bibitem{Sadhasivam2023}
V. G. Sadhasivam, L. Meuser, D. R. Reichman, and S. C. Althorpe, ``Instantons 
and the quantum bound to chaos,'' \textit{Proc. Natl. Acad. Sci. U.S.A.}, 
vol. 120, e2312378120, 2023. 

\bibitem{Wolynes2023}
P. G. Wolynes and M. Gruebele, ``Quantum scrambling across an energy barrier,'' 
\textit{Proc. Natl. Acad. Sci. U.S.A.}, vol. 120, e2319705120, 2023.

\bibitem{Sadhasivam2025}
V. G. Sadhasivam, J. M. Rost, and S. C. Althorpe, ``On the origin of 
exponential operator growth in Hilbert space,'' \textit{arXiv:2511.02800}, 2025. 

\bibitem{Miller1975}
W. H. Miller, ``Semiclassical limit of quantum mechanical transition state theory 
for nonseparable systems,'' \textit{J. Chem. Phys.}, vol. 62, pp. 1899-1906, 1975.

\bibitem{Gutzwiller1971}
M. C. Gutzwiller, ``Periodic Orbits and Classical Quantization Conditions,'' 
\textit{J. Math. Phys.}, vol. 12, no. 3, pp. 343-358, 1971. 

\bibitem{Berry1976}
M. V. Berry and M. Tabor, ``Closed orbits and the regular bound spectrum,'' 
\textit{Proc. R. Soc. Lond. A}, vol. 349, no. 1656, pp. 101-123, 1976. 

\bibitem{Richter2022}
K. Richter, J. D. Urbina, and S. Tomsovic, ``Semiclassical roots of 
universality in many-body quantum chaos,'' \textit{J. Phys. A: Math. Theor.}, 
vol. 55, 453001, 2022. 

\bibitem{Schubert2010}
R. Schubert, H. Waalkens, A. Goussev, and S. Wiggins, ``Periodic-orbit 
formula for quantum reactions through transition states,'' \textit{Phys. Rev. 
A}, vol. 82, 012707, 2010. 

\bibitem{Combescure1999} 
M. Combescure, J. Ralston, and D. Robert, ``A proof of the Gutzwiller 
semiclassical trace formula using coherent states decomposition,'' 
\textit{Commun. Math. Phys.}, vol. 202, pp. 463-480, 1999.

\bibitem{RomeroBermudez2019}
A. Romero-Bermúdez, K. Schalm, and V. Scopelliti, ``Regularization dependence 
of the OTOC. Which Lyapunov spectrum is the physical one?,'' \textit{JHEP} 
07 (2019) 107, 2019.

\bibitem{Bouzouina2002}
A. Bouzouina and D. Robert, ``Uniform semiclassical estimates for the propagation of quantum observables,'' \textit{Duke Math. J.}, vol. 111, no. 2, pp. 223-252, 2002.

\bibitem{Trunin2023}
D. A. Trunin, ``Out-of-time-order correlators and quantum chaos,'' \textit{Phys. 
Usp.} \textbf{66}, 1176 (2023); arXiv:2309.14480.

\bibitem{Trunin2023False}
D. A. Trunin, ``Quantum chaos without false positives,'' \textit{Phys. Rev. D}, 
vol. 108, L101703, 2023.

\bibitem{Trunin2023Replica}
D. A. Trunin, ``Refined quantum Lyapunov exponents from replica out-of-time-order 
correlators,'' \textit{Phys. Rev. D}, vol. 108, 105023, 2023.

\bibitem{Jalabert2001}
R. A. Jalabert and H. M. Pastawski, ``Environment-independent decoherence 
rate in classically chaotic systems,'' \textit{Phys. Rev. Lett.}, vol. 86, 
2490, 2001.

\bibitem{Goussev2012}
A. Goussev, R. A. Jalabert, H. M. Pastawski, and D. Wisniacki, ``Loschmidt 
Echo,'' \textit{Scholarpedia}, vol. 7, no. 8, p. 11687, 2012.

\bibitem{Zhang2005}
F. Zhang, \textit{The Schur Complement and Its Applications}, Springer, 2005.

\bibitem{Wiggins1994}
S. Wiggins, \textit{Normally Hyperbolic Invariant Manifolds in Dynamical 
Systems}, Springer, 1994.

\bibitem{Waalkens2008}
H. Waalkens, R. Schubert, and S. Wiggins, ``Wigner's dynamical transition 
state theory in phase space: classical and quantum,'' \textit{Nonlinearity}, 
vol. 21, R1, 2008. 

\end{thebibliography}
\end{document}